\documentclass[aps,prl,twocolumn,superscriptaddress,showpacs,amsmath,floatfix]
{revtex4}
\usepackage{graphicx}
\usepackage{bm}

\begin{document}

\title{Selective population and neutron decay of the first excited state of 
semi-magic ${\mathrm{^{23}}}$O}

\author{A.~Schiller}
\email{schiller@nscl.msu.edu}
\affiliation{National Superconducting Cyclotron Laboratory, Michigan State 
University, East Lansing, MI 48824}
\author{N.~Frank}
\affiliation{National Superconducting Cyclotron Laboratory, Michigan State 
University, East Lansing, MI 48824}
\affiliation{Department of Physics \& Astronomy, Michigan State University, 
East Lansing, MI 48824}
\affiliation{Department of Physics, Concordia College, Moorhead, MN 56562}
\author{T.~Baumann}
\affiliation{National Superconducting Cyclotron Laboratory, Michigan State 
University, East Lansing, MI 48824}
\author{D.~Bazin}
\affiliation{National Superconducting Cyclotron Laboratory, Michigan State 
University, East Lansing, MI 48824}
\author{B.A.~Brown}
\affiliation{National Superconducting Cyclotron Laboratory, Michigan State 
University, East Lansing, MI 48824}
\affiliation{Department of Physics \& Astronomy, Michigan State University, 
East Lansing, MI 48824}
\author{J.~Brown}
\affiliation{Department of Physics, Wabash College, Crawfordsville, IN 47933}
\author{P.A.~DeYoung}
\affiliation{Department of Physics, Hope College, Holland, MI 49423}
\author{J.E.~Finck}
\affiliation{Department of Physics, Central Michigan University, Mt.\ Pleasant,
MI 48859}
\author{A.~Gade}
\affiliation{National Superconducting Cyclotron Laboratory, Michigan State 
University, East Lansing, MI 48824}
\affiliation{Department of Physics \& Astronomy, Michigan State University, 
East Lansing, MI 48824}
\author{J.~Hinnefeld}
\affiliation{Department of Physics \& Astronomy, Indiana University at South 
Bend, South Bend, IN 46634}
\author{R.~Howes}
\affiliation{Department of Physics, Marquette University, Milwaukee, WI 53201}
\author{J.-L.~Lecouey}
\altaffiliation[Present address: ]{Laboratoire de Physique Corpusculaire, 
ENSICAEN, IN2P3, 14050 Caen, Cedex, France}
\affiliation{National Superconducting Cyclotron Laboratory, Michigan State 
University, East Lansing, MI 48824}
\author{B.~Luther}
\affiliation{Department of Physics, Concordia College, Moorhead, MN 56562}
\author{W.A.~Peters}
\affiliation{National Superconducting Cyclotron Laboratory, Michigan State 
University, East Lansing, MI 48824}
\affiliation{Department of Physics \& Astronomy, Michigan State University, 
East Lansing, MI 48824}
\author{H.~Scheit}
\affiliation{National Superconducting Cyclotron Laboratory, Michigan State 
University, East Lansing, MI 48824}
\author{M.~Thoennessen}
\affiliation{National Superconducting Cyclotron Laboratory, Michigan State 
University, East Lansing, MI 48824}
\affiliation{Department of Physics \& Astronomy, Michigan State University, 
East Lansing, MI 48824}
\author{J.A.~Tostevin}
\affiliation{National Superconducting Cyclotron Laboratory, Michigan State 
University, East Lansing, MI 48824}
\affiliation{Department of Physics \& Astronomy, Michigan State University, 
East Lansing, MI 48824}
\affiliation{School of Electronics and Physical Sciences, University of Surrey,
Guildford GU2 7XH, UK}

\begin{abstract}
We have observed an excited state in the neutron-rich semi-magic nucleus 
$^{23}$O for the first time. No such states have been found in previous 
searches using $\gamma$-ray spectroscopy. The observation of a resonance in 
$n$-fragment coincidence measurements confirms the speculation in the 
literature that the lowest excited state is neutron unbound and establishes 
positive evidence for a 2.8(1)~MeV excitation energy of the first excited state
in $^{23}$O\@. The non-observation of a predicted second excited state is 
explained assuming selectivity of inner-shell knockout reactions.
\end{abstract}

\pacs{21.10.Pc, 23.90.+w, 25.60.Gc, 29.30.Hs}

\maketitle

Magic numbers are pillars of our understanding of nuclear structure. For each 
magic number of nucleons one can observe, e.g., (i) a large number of stable 
isotopes or isotones, large binding energies, and corresponding structures in 
separation-energy systematics, (ii) increased excitation energies of first 
excited states for even-even nuclei, and a corresponding reduction of their 
ground-state transition matrix elements, and (iii) a decreased level density at
low and moderate excitation energies \cite{BM69}\@. Nuclei with one nucleon 
added or removed from a doubly-magic core are well described by a 
single-particle (SP) model; their properties such as spins and parities, 
magnetic moments, energies of the lowest excited states, and transition matrix 
elements are well reproduced by SP estimates. The underlying physics for the 
emergence of magic numbers is the presence of gaps in the SP spectrum around 
the Fermi energy, which makes the last nucleon tightly bound whereas any 
additional nucleons will be loosely bound. If the energy gaps are larger than 
the strength of residual interactions such as pairing or quadrupole-quadrupole 
interactions, only weak correlations of such types are induced, hence, no 
corresponding low-lying collectivity is observed. 

Magic numbers for stable nuclei follow harmonic-oscillator shells up to 
$N,Z=20$; for heavier nuclei, the spin-orbit splitting between high-$l$ 
orbitals produces $N,Z=28$, 50, 82, and 126\@. The emergence of new magic 
numbers far from stability is thought to be related to the effect of the tensor
force \cite{OS05+OM06}, which produces (among others) attractive contributions 
for $0d_{3/2}$ neutrons in the presence of $0d_{5/2}$ protons. For oxygen 
isotopes the proton $0d_{5/2}$ orbit is not occupied. This contributes to the 
reduced binding of the $\nu(0d_{3/2})$ orbital \cite{OF01} and produces a gap 
at $N=16$, while for stable nuclei the gap at $N=16$ disappears since the 
$0d_{5/2}$ proton shell is almost filled. The effect of the occupation number 
on SP spectra is taken into account by considering effective SP (ESP) energies,
where SP energies are modified by the average monopole contribution of residual
interactions. The neutron ESP spectra for neutron-rich oxygen isotopes derived 
with the universal $sd$-shell (USD) interaction (which takes into account the 
effects of residual forces by virtue of a fit to experimental data) reveal 
large gaps for $^{22}$O and $^{24}$O at the Fermi energy. Both nuclei are 
predicted to be doubly magic with $N=14$ and 16, respectively \cite{BR05}; 
hence, the properties of $^{23}$O are expected to be well reproduced by a SP 
description. In $^{22}$O, a high-lying first excited state has been observed in
$\gamma$-ray spectroscopy \cite{SA04} confirming the magic character of $N=14$ 
for $Z=8$\@. Systematics of neutron separation energies suggest a magic 
character of $N=16$ at the neutron dripline \cite{OK00}, but no high-lying 
first excited state has been found for $^{24}$O \cite{SA04}\@. Focusing on the 
SP nucleus $^{23}$O, we have populated the $n$-$^{22}$O continuum by $2p+1n$ 
removal from a $^{26}$Ne beam and deduced the decay-energy spectrum of excited 
$^{23}$O in order to search for resonances corresponding to unbound states. 
Two-proton knockout reactions have proven to be a sensitive tool to study 
nuclear structure of rare isotopes \cite{BB03+Wa03+FW05}\@. These direct 
reactions rely on the knockout of valence protons leaving the remaining 
nucleons largely undisturbed. In the present paper, direct knockout of 
inner-shell protons is utilized for the first time to observe neutron-unbound 
states in neutron-rich nuclei. It is shown that these proton hole states couple
to specific highly excited neutron states.

\begin{figure}[t!]
\includegraphics[totalheight=3.7cm]{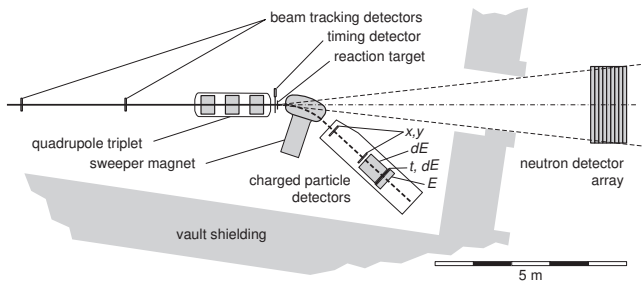}
\caption{Experimental setup for measuring fragment-neutron coincidences at the 
NSCL\@. The A1900 and extended focal plane scintillator are upstream and not 
shown here.}
\label{fig:setup}
\end{figure}

The experiment was performed at the National Superconducting Cyclotron 
Laboratory (NSCL) at Michigan State University. A 105~pnA primary beam of 
140~MeV/$A$ $^{40}$Ar impinged on a 893~mg/cm$^2$ Be production target. The 
resulting cocktail beam was purified with respect to the desired $^{26}$Ne at 
86~MeV/$A$ using an achromatic 750-mg/cm$^2$-thick acrylic wedge degrader and 
either 1\% or 3\% momentum-acceptance slits at the dispersive focal plane for 
different parts of the experiment, as well as $\pm$10~mm slits at the extended 
focal plane of the A1900 fragment separator \cite{MS03}\@. A purity of up to 
93.2\% was achieved with a $^{26}$Ne beam intensity of about 7000~pps. The 
contaminants (mainly $^{27}$Na and $^{29}$Mg) were separated event-by-event in 
the off-line analysis by their different time-of-flight (ToF) from the extended
focal plane to the scintillator in front of the 721-mg/cm$^2$-thick Be reaction
target (see Fig.\ \ref{fig:setup})\@. Positions and angles of incoming beam 
particles were measured by two $15\times 15$~cm$^2$ position-sensitive 
parallel-plate avalanche counters (PPACs) with 20~pads/inch, i.e., 
${\mathrm{FWHM}}\approx 1.3$~mm. Due to the focusing of the quadrupole triplet,
positions of impinging $^{26}$Ne particles on the reaction target could be 
reconstructed within a radius of $\sigma_r=1$~mm. 

\begin{figure}[t!]
\includegraphics[totalheight=8.6cm]{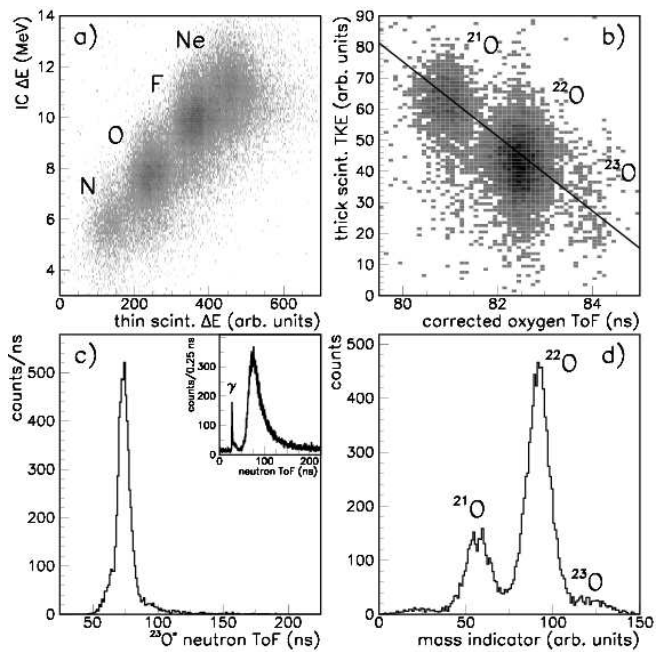}
\caption{a) Energy loss in thin scintillator vs.\ ion chamber provides $Z$ 
identification. b) TKE in thick scintillator vs.\ corrected ToF for oxygen 
fragments provides isotope identification. c) Neutron ToF for decay of 
$^{23}$O$^*$, i.e., in coincidence with $^{22}$O fragments, and using a thick 
target in singles mode showing prompt $\gamma$ radiation (inset)\@. The 1.4~ns 
FWHM of the $\gamma$ peak translates into a $n$-energy resolution of 1.4~MeV\@.
d) Data in b) projected onto the diagonal line.}
\label{fig:id}
\end{figure}

Charged particles behind the reaction target were bent $43^\circ$ by the 
large-gap 4~Tm Sweeper Magnet \cite{BK05}\@. Two $30\times 30$~cm$^2$ 
cathode-readout drift chambers (CRDCs) provided position in the dispersive 
(10~pads/inch, $\sigma_x\approx 1.1$~mm) and non-dispersive (drift time, 
$\sigma_y\approx 1.3$~mm) direction. The 1.87~m distance between the CRDCs 
translates this into $\sigma=1$~mrad angle resolution. Energy loss was 
determined in a 65-cm-long ion chamber (IC) and a $40\times 40$~cm$^2$, 
4.5-mm-thin plastic scintillator whose pulse-height signal was corrected for 
position. Energy loss was used to separate reaction products with different $Z$
(see Fig.\ \ref{fig:id}a)\@. The thin scintillator also gave ToF of reaction 
products from the reaction target. This, together with the total kinetic energy
(TKE) measurement in a 15-cm-thick plastic scintillator, provided isotopic 
separation (see Fig.\ \ref{fig:id}b, d)\@. The pulse-height signal of the thick
scintillator was also corrected for position; the raw ToF was corrected for (i)
position on the thin scintillator, for (ii) angle and position behind the 
Sweeper Magnet, as well as for (iii) the point of interaction on the target. 
The ToF corrections approximately compensate the spread in energy of reaction 
products and account for differences in length of the fragment tracks and of 
the light paths in the thin scintillator. Due to the finite acceptance a 
magnetic-rigidity range of reaction products is selected. For reaction products
with equal $Z$, this translates into $\mathrm{TKE}\propto 1/A$ and 
$\mathrm{ToF}\propto A$ (see Fig.\ \ref{fig:id}b)\@. 

Neutrons were detected in the Modular Neutron Array (MoNA) \cite{LB03+BB05} at 
a distance of 8.2~m from the reaction target. MoNA consists of $9\times 16$ 
stacked 2-m-long plastic scintillator bars which are read out on both ends by 
photomultiplier tubes (PMT)\@. The bars are mounted horizontally and 
perpendicular to the beam axis. Position along the vertical and along the beam 
axis is determined within the thickness of one bar (10~cm, 
$\sigma\sim 3$~cm)\@. Horizontal position and neutron ToF are determined by the
time difference and the mean time, respectively, of the two PMT signals which 
yield resolutions of $\sigma\approx 5$~cm and $\sigma\approx 0.1$~ns. The ToF 
spectrum was calibrated by shifting the prompt $\gamma$-ray peak from a 
thick-target MoNA singles run to 27~ns, see inset in Fig.\ \ref{fig:id}c. A 
neutron ToF spectrum (first hit after $\sim 27$~ns) in coincidence with 
$^{22}$O fragments is given in Fig.\ \ref{fig:id}c. 

The decay energy of resonances is reconstructed by the invariant mass method. 
For this, the relativistic four-momentum vectors of the neutron and fragment 
are reconstructed at the point of breakup. For neutrons, position and ToF 
resolution translate into angle and energy resolution of $\sigma=8$~mrad and 
$\sigma=1.6$~MeV \footnote{The nominal energy resolution of $\sigma=0.7$~MeV is
degraded by a systematic ToF walk with deposited energy. This walk does neither
affect the position resolution along a bar, nor can it be corrected due to 
inefficiencies of the charge-signal collection. Furthermore, the effect on the 
decay energy resolution is negligible as it is determined mostly by the neutron
angle resolution and the target thickness.}, respectively. The angle and energy
of fragments in coincidence with neutrons were reconstructed behind the 
reaction target based on the ion-optical properties of the Sweeper Magnet using
a novel method which takes into account the position at the target in the 
dispersive direction \cite{Fr06}\@. The reconstructed position in the 
non-dispersive direction serves thereby as a double check on an event-by-event 
basis against the same information obtained from forward tracking from the 
PPACs through the quadrupole triplet. The angle and energy resolution of the 
fragments behind the target are $\sigma=2.7$~mrad and $\sigma=0.4$~MeV/$A$, 
respectively. The average energy loss of the fragment through half of the 
reaction target is added to approximate the relativistic four-momentum vector 
of the fragment at the average breakup point. 

\begin{figure}[t!]
\includegraphics[totalheight=4.3cm]{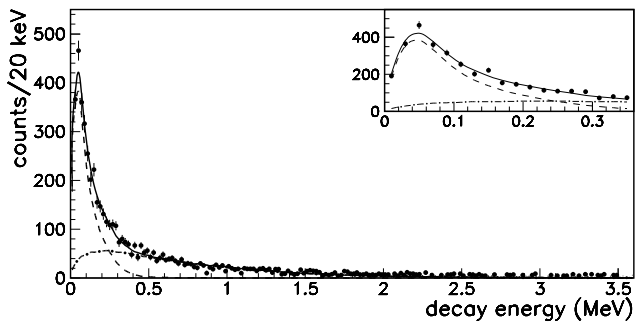}
\caption{Decay-energy spectra of $^{23}$O$^*$\@. The inset shows a close-up of 
the first 360~keV\@. Smooth curves correspond to simulations (see text)\@. The 
solid curve corresponds to the sum of the dash-dotted and dashed curves.}
\label{fig:decay}
\end{figure}

The experiment ran for two days and produced 5700 $n$-$^{22}$O coincidences. 
The reconstructed decay-energy spectrum from $n$-$^{22}$O coincidences is shown
in Fig.\ \ref{fig:decay}\@. This spectrum is affected by the finite acceptance 
for fragments and neutrons and a bias toward the fastest neutron for events 
where more than one neutron hit MoNA\@. The latter is because in the analysis 
only the first hit after $\sim 27$~ns is taken into account. Later hits due to 
either multiple scattering of one neutron or true multiple-neutron events are 
ignored in the analysis \cite{Fr06}\@. The most severe of these effects, namely
the acceptance cuts, have been simulated. For the simulation, a Glauber 
reaction model is used. Angle straggling of the fragments in the target is 
taken into account, as well as detector resolutions. The $n$-$^{22}$O data in 
Fig.\ \ref{fig:decay} are best described by a Breit-Wigner resonance at 
45(2)~keV decay energy (dashed curve) on top of a beam-velocity source of 
Maxwellian-distributed neutrons (a thermal model) with $T\sim 0.7$~MeV 
(dash-dotted curve)\@. The ratio of the contributions is 1:2\@. The simulated 
decay-energy resolution, in qualitative agreement with \cite{FN04}, is found to
scale as ${\mathrm{FWHM}}\sim 16\sqrt{E\cdot{\mathrm{keV}}}$ and 
$40\sqrt{E\cdot{\mathrm{keV}}}$, below and above $E=1.5$~MeV decay energy, 
respectively. The numerical factors are dominated by $n$-angle resolution and 
target thickness, respectively. For the present resonance, the width due to 
experimental conditions (100~keV) overshadows the Wigner limit by about three 
orders of magnitude, hence its lifetime is not determined. The thermal model 
has been included in the fit because it is a good description of uncorrelated 
events as determined from event mixing, and it serves as an estimate of 
non-resonant contributions to the decay-energy spectrum (see, e.g., 
\cite{FN04,DK87} for further discussions of experimental backgrounds)\@. 

\begin{figure}[t!]
\includegraphics[totalheight=3.4cm]{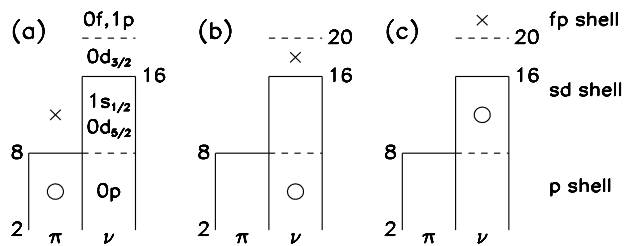}
\caption{a) Likely configuration of the $^{24}$O continuum populated by direct 
$2p$ removal from $^{26}$Ne. These negative-parity proton excitations will mix 
with certain negative-parity neutron excitations b), c), which then decay into 
the $^{23}$O ground state or continuum.}
\label{fig:levels}
\end{figure}

When combined with the neutron separation energy $S_n$ \cite{AW03}, our result 
establishes an excitation energy of 2.79(13)~MeV for the first excited state of
$^{23}$O\@. Two low-lying states with spin-parities of $3/2^+$ and $5/2^+$, 
based on a $\nu(0d_{3/2})^1$ particle and $\nu(0d_{5/2})^{-1}$ hole 
configuration, respectively, are predicted by theory \cite{VZ05}, with the 
$5/2^+$ level the lower one and hence being the favored spin-parity assignment 
for the here observed resonance. Observation of decay to excited states in 
$^{22}$O which are at 3.4~MeV and higher are thought to be unlikely.

The selective population of the states can be understood from the structure of 
$^{24}$O and the specific reaction mechanism. The main component (90\%) of the 
$^{24}$O ground-state wavefunction has the closed-shell configuration 
$[d_{5/2}^6,s_{1/2}^2]$\@. The $^{26}$Ne ground-state wavefunction is dominated
by two $sd$-shell protons coupled to the ground state of $^{24}$O 
\cite{TB06}\@. Two-proton removal from $^{26}$Ne to the low-lying states in 
$^{24}$O can thus proceed by the removal of two $sd$-shell protons to the 
$^{24}$O ground state or by one $sd$-shell and one $p$-shell proton to the 
negative-parity configuration shown in Fig.\ \ref{fig:levels}a. The 
negative-parity eigenstates are linear combinations of the three configurations
in Fig.\ \ref{fig:levels}\@. The neutron-decay of these eigenstates goes by 
$d_{3/2}$ neutron decay to high-lying negative parity states (b) and by 
$fp$-shell neutron decay to the $1/2^+$ and $5/2^+$ one-neutron-hole states (c)
in $^{23}$O\@. Thus, starting with the main component of the $^{26}$Ne 
wavefunction no overlap with the $3/2^+$ $[d_{5/2}^6,d_{3/2}]$ excited state of
$^{23}$O is possible. The direct reaction theory \cite{To06} for the removal of
two $sd$-shell protons from the $^{26}$Ne to the $^{24}$O ground state gives a 
cross section of about 0.42~mb where a suppression factor of 0.5 was used 
\cite{To06}\@. The removal of one proton from the $sd$-shell and one proton 
from the fully occupied $p$-shell (6 protons) corresponds to a total cross 
section of about 2.1~mb. These states will neutron decay to the low-lying 
$1/2^+$ and $5/2^+$ states of $^{23}$O\@. In addition, we might expect some 
direct three-nucleon knockout from removal of two $sd$-shell protons plus one 
$sd$-shell neutron, but again, with the dominant $[d_{5/2}^6,s_{1/2}^2]$ 
neutron structure of $^{26}$Ne this can only go to the 
$1/2^+$ $[d_{5/2}^6,s_{1/2}^1]$ and $5/2^+$ $[d_{5/2}^5,s_{1/2}^2]$ states of 
$^{23}$O\@. Population of the $3/2^+$ $[d_{5/2}^6,d_{3/2}]$ state of $^{23}$O 
as well as the $2^+$ $[d_{5/2}^6,s_{1/2},d_{3/2}]$ state of $^{24}$O can only 
come from the small component of the $^{26}$Ne ground state with the neutron 
configuration $[d_{5/2}^6,s_{1/2},d_{3/2}]\otimes 2^+(\pi)$\@. With full 
$sd$-shell wavefunctions the cross section to the $2^+$ state of $^{24}$O is 
0.03~mb (with the reduction factor of 0.5) compared to 0.42~mb for the ground 
state. Accordingly, the cross section for the $3/2^+$ state in $^{23}$O should 
be smaller by about an order of magnitude than those for the $1/2^+$ ground and
$5/2^+$ excited states. 

Considering only the dominant components of the wavefunctions the spectroscopic
factor for the neutron decay of the $5/2^+$ $[d_{5/2}^5,s_{1/2}^2]$ state in 
$^{23}$O to the $[d_{5/2}^6]$ ground state in $^{22}$O is zero. With the full 
USDB wavefunctions \cite{BR06}, the spectroscopic factor is 0.059\@. Thus, with
a $d$-wave neutron single-particle width of 90(10)~eV (corresponding to a decay
energy of 45(2)~keV), the total decay width of 5.0(6)~eV is extremely small. 
Still, the calculated $\gamma$-decay lifetime is 4.5~ps corresponding to a 
partial width of 0.15~meV, hence the decay is dominated by neutrons. 

In conclusion, we have observed for the first time an excited state in 
$^{23}$O\@. This state has been sought for unsuccessfully using $\gamma$-ray 
spectroscopy; due to its unbound nature it is revealed in the present work in 
an $n$-fragment coincidence experiment. On the other hand, a predicted second 
excited state has not been found. This has been explained by the presence of 
selection effects in the population of states in $^{23}$O, which, in return, 
demonstrates that inner-shell proton knockout is a new promising spectroscopic 
tool to explore excited states far from stability.

We would like to thank the members of the MoNA collaboration G. Christian, C. 
Hoffman, K.L. Jones, K.W. Kemper, P. Pancella, G. Peaslee, W. Rogers, S. Tabor,
and about 50 undergraduate students for their contributions to this work. We 
would like to thank R.A. Kryger, C. Simenel, J.R. Terry, and K. Yoneda for 
their valuable help during the experiment. Financial support from the National 
Science Foundation under grant numbers PHY-01-10253, PHY-03-54920, 
PHY-05-55366, PHY-05-55445, and PHY-06-06007 is gratefully acknowledged. J.E.F.
and J.T. acknowledge support from the Research Excellence Fund of Michigan and 
from the United Kingdom Engineering and Physical Sciences Research Council 
(EPSRC) under Grant No.\ EP/D003628, respectively.


\begin{thebibliography}{}
\bibitem{BM69}Aa.\ Bohr and B.R. Mottelson, \em Nuclear Structure \rm 
(Benjamin, New York, 1969), Vol.\ I, p.\ 189.
\bibitem{OS05+OM06}T. Otsuka, T. Suzuki, R. Fujimoto, H. Grawe, and Y. Akaishi,
Phys.\ Rev.\ Lett.\ \bf 95\rm, 232502 (2005); T. Otsuka, T. Matsuo, and D. Abe,
\it ibid.\ \bf 97\rm, 162501 (2006).
\bibitem{OF01}T. Otsuka \sl et al.\rm, Phys.\ Rev.\ Lett.\ \bf 87\rm, 082502 
(2001).
\bibitem{BR05}B.A. Brown and W.A. Richter, Phys.\ Rev.\ C \bf 72\rm, 057301 
(2005).
\bibitem{SA04}M. Stanoiu \sl et al.\rm, Phys.\ Rev.\ C \bf 69\rm, 034312 
(2004).
\bibitem{OK00}A. Ozawa, T. Kobayashi, T. Suzuki, K. Yoshida, and I. Tanihata, 
Phys.\ Rev.\ Lett.\ \bf 84\rm, 5493 (2000).
\bibitem{BB03+Wa03+FW05}D. Bazin \it et al.\rm, Phys.\ Rev.\ Lett.\ \bf 91\rm, 
012501 (2003); D. Warner, Nature \bf 425\rm, 570 (2003); J. Fridmann \it et 
al.\rm, \it ibid.\ \bf 435\rm, 922 (2005).
\bibitem{MS03}D.J. Morrissey, B.M. Sherrill, M. Steiner, A. Stolz, and I. 
Wiedenhoever, Nucl.\ Instrum.\ Methods Phys.\ Res. \bf B204\rm, 90 (2003).
\bibitem{BK05}M.D. Bird \sl et al.\rm, IEEE Trans.\ Appl.\ Supercond.\ \bf 
15\rm, 1252 (2005).
\bibitem{LB03+BB05}B. Luther \sl et al.\rm, Nucl.\ Instrum.\ Methods Phys.\ 
Res.\ \bf A505\rm, 33 (2003); T. Baumann \sl et al.\rm, \it ibid.\ \bf A543\rm,
517 (2005).
\bibitem{Fr06}N. Frank, Ph.D. thesis, Michigan State University, 2006.
\bibitem{FN04}N. Fukuda \sl et al.\rm, Phys.\ Rev.\ C \bf 70\rm, 054606 (2004).
\bibitem{DK87}F. De\'{a}k, \sl et al.\rm, Nucl.\ Instrum.\ Methods Phys.\ Res.\
\bf A258\rm, 67 (1987).
\bibitem{AW03}G. Audi, A.H. Wapstra, and C. Thibault, Nucl.\ Phys.\ \bf 
A729\rm, 337 (2003).
\bibitem{VZ05}A. Volya and V. Zelevinsky, Phys.\ Rev.\ Lett.\ \bf 94\rm, 052501
(2005).
\bibitem{TB06}J.R. Terry \sl et al.\rm, Phys.\ Lett.\ \bf B 640\rm, 86 (2006).
\bibitem{To06}J.A. Tostevin and B.A. Brown \rm, Phys.\ Rev.\ C \bf 74\rm, 
064604 (2006).
\bibitem{BR06}B.A. Brown and W.A. Richter \rm, Phys.\ Rev.\ C \bf 74\rm, 034313
(2006).
\end{thebibliography}
\end{document}